\documentclass[11pt,twoside]{article}
\usepackage{graphicx}
\usepackage{wrapfig}
\usepackage{asp2006}
\usepackage{lscape}
\def\edcomment#1{\iffalse\marginpar{\raggedright\sl#1\/}\else\relax\fi}
\reversemarginpar

\begin{document}
\title{Molecular gas in  nearby elliptical radio galaxies.}
\author{B. Oca\~na Flaquer and S. Leon}
\affil{Instituto de Radio Astronom\'ia Milim\'etrica - IRAM - Av. Divina Pastora 7, N\'ucleo Central, 18012 Granada, Spain.}
\author{L. Lim and Dinh-V-Trung }
\affil{Institute of Astronomy and Astrophysics, Academia Sinica, 128, Sec. 2, Yen Gao Yuan Road, Nankang, Taipei, Taiwan, R.O.C.}
\author{F. Combes}
\affil{Observatoire de Paris, LERMA, 61 Av. del Observatoire, F-75014. Paris, France}

\begin{abstract}
Powerful radio-AGN  are normally  hosted by massive elliptical galaxies which are usually very poor in molecular gas. Nevertheless the gas is needed in the very center to feed the nuclear activity. Thus it is important to study the origin, the distribution and the kinematics of the molecular gas in such objects. We have performed at the IRAM-30m telescope a  survey of the CO(1-0) and CO(2-1) emission in the most powerful radio galaxies of the Local Universe, selected only on the basis of their radio-continuum fluxes. The main result of this survey is the very low content in molecular gas of such galaxies compared to FIR selected galaxies. The median value of the molecular gas mass, taking into account the upper limits,  is $1\times10^8~M_{\odot}$; if we calculate it for all the galaxies together, and if we separate them into FR-I and FR-II type galaxies, an important difference is found between them. Moreover, the CO spectra indicate the presence of a central molecular gas disk in these radio galaxies. Our results contrast with  previous surveys, mainly selected through the FIR emission, with a larger mass of molecular gas observed. The first results indicate that minor mergers are good candidates to fuel the central part of the radio galaxies of our sample.
\end{abstract}

\vspace{-0.5cm}
\section{Introduccion}
Radio Galaxies are identified with strong radio sources in the range of $10^{41}$ to $10^{46}erg/s$. Most of them are giant elliptical galaxies that contain little dense and cold interstellar medium (ISM) (\cite{Kellerman88}). They have been detected in the far infra red (FIR), which is assumed to be thermal and to originate from dust heated by young massive stars or an AGN (\cite{Wiklind95b}). The powerful radio-AGNs are usually very poor in molecular gas, nevertheless the central black hole (BH) needs gas to feed the nuclear activity and this is why we need to study the origin, the distribution and kinematics of the molecular gas in such objects. \cite{Antonucci93} suggests that AGNs could be powered by accretion of ISM into super massive black holes (SMBHs), and according to \cite{deRuiter02} the presence of ISM in the circumnuclear region of AGNs is indeed inevitable and the large scale dust/gas system are related to nuclear activity.\\


\section{Sample and observations}
We have a total of 52 nearby radio galaxies in our sample, chosen on the basis of their radio continuum power. The selection criteria of this sample makes it different from other samples, \cite{Evans05,Mazzarella93,Bertram07}, which are chosen by their IR fluxes or because they show signs of interactions. The galaxies were observed using the IRAM-30m telescope for CO(1-0) and CO(2-1) at 115 and 230 GHz respectively.\\ 
Out of the 52 galaxies, 43 have 12CO(1-0) and 12CO(2-1) data and 9 of them only have 12CO(1-0) data. From the total 52 galaxies we have  55\% of the galaxies detected either on the 12CO(1-0) line, the 12CO(2-1) line or in both lines together; note that this 55\% includes galaxies tentatively detected as well as detected. From the clearly detected galaxies we do have a 38\% detection rate in either one of the transition lines. 7 of our galaxies have been detected in both transition lines and, except for one of them, all have a higher  integrated emission in the 12CO(2-1) line than in the 12CO(1-0) line. 90\% of the 52 were detected in the continuum at 3mm, and 55\% of the 43 were detected in the continuum at 1mm.\\

\section{Molecular gas mass}

\begin{wrapfigure}[19]{r}{80mm} 
\centering
\includegraphics[scale=0.30]{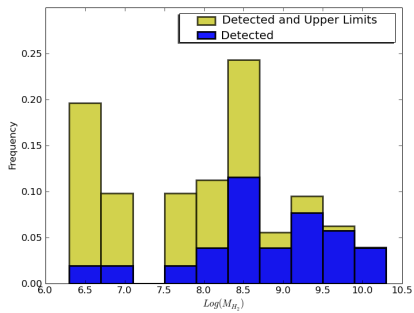}
\caption{$M_{H_2}$ histogram of all the galaxies calculated using ASURV. On the front, the darker color represents the part of the histogram that belongs to the detected galaxies only, leaving the rest of the histogram to the upper limits exclusively.}
\label{mh2histo}
\end{wrapfigure}

The total H$_2$ mass was calculated using the standard CO-to-H$_2$ conversion factor of $2.3\times10^{20} cm^{-2} (K  km s^{-1})^{-1}$ with an average gas mass of $1\times10^8~ M_{\odot}$. This average was calculated using the survival analysis statistics (ASURV) which takes into account the upper limits. We compared the molecular gas mass average of our sample with the molecular gas mass of other samples. We noticed that \cite{Wiklind95a}, with a sample of elliptical galaxies, has an average molecular gas mass of the same order of magnitude as our sample. We also compared with \cite{Evans05,Mazzarella93,Bertram07}. Their sample is a FIR selected sample or galaxies in interaction, and their molecular gas masses are in the range of $10^9$. Finally, we also compared with \cite{Solomon97}, with a sample of ULIRGS which has an average of $10^{10}~ M_{\odot}$ of molecular gas.

\subsection{Fanaroff and Riley classification}
\begin{figure}
\begin{center}
\begin{tabular}{cc}
\includegraphics[scale=0.25]{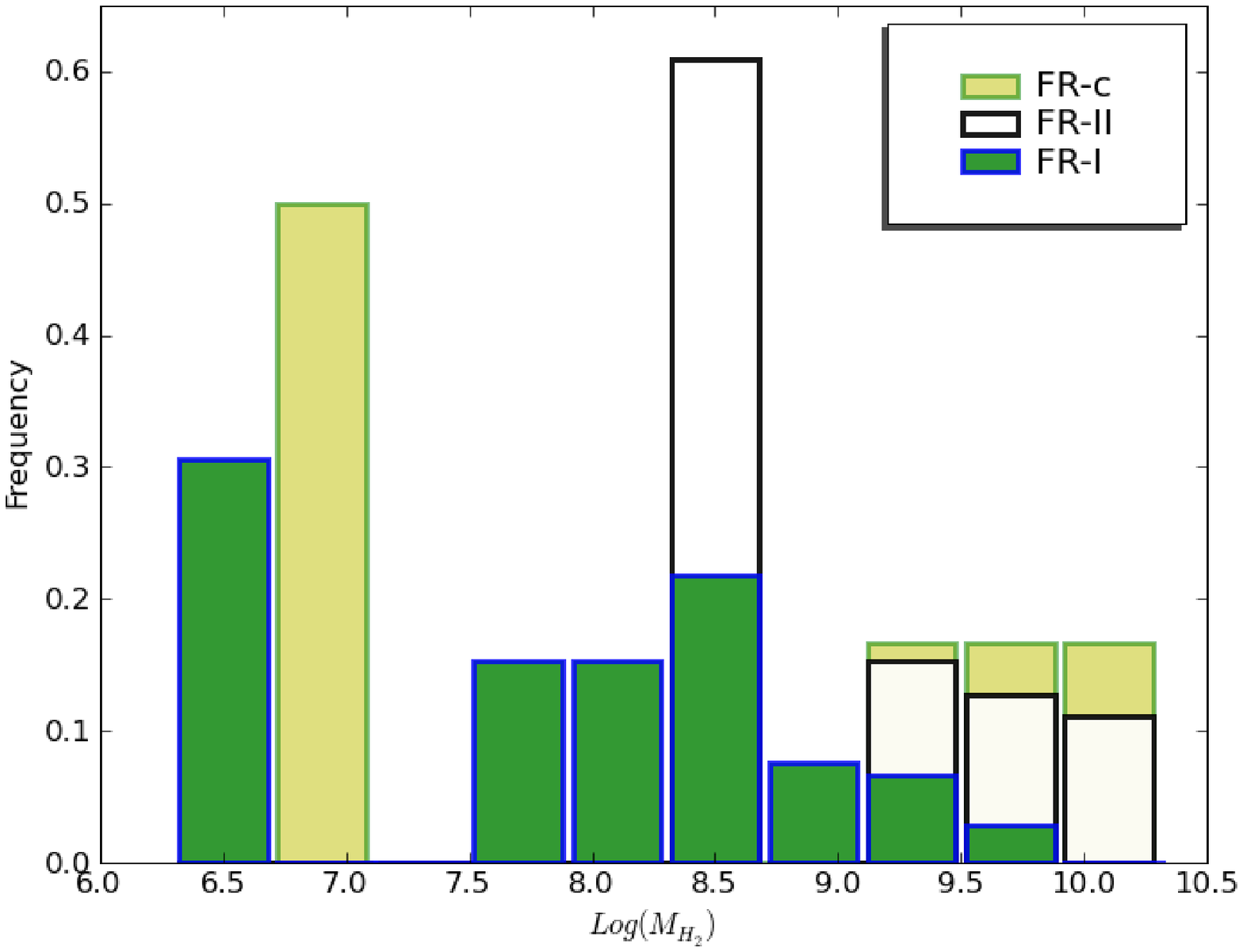} & \includegraphics[scale=0.3]{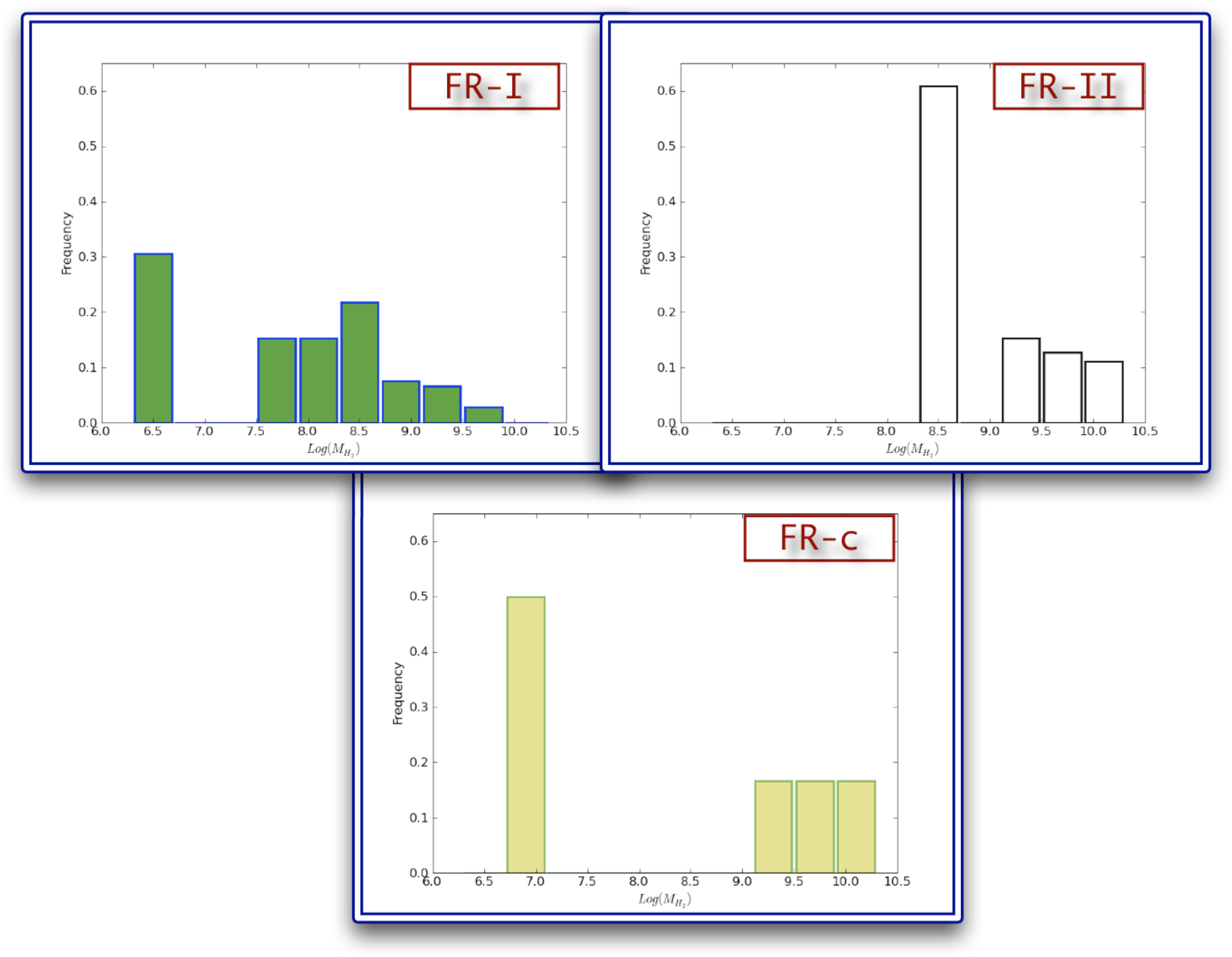}
\end{tabular}
\caption{$M_{H_2}$ distribution of the galaxies depending on their Fanaroff-Riley classification. On the left  we represent a histogram with all of the galaxies together, with the white color for the FR-II type, the darker color for the FR-I and the brighter color for the FR-c; and on the right we can see the same classification but shown separately for each type.}
\label{FRhisto}
\end{center}
\end{figure}
\begin{wrapfigure}[16]{r}{75mm}
\centering
\includegraphics[scale=0.3]{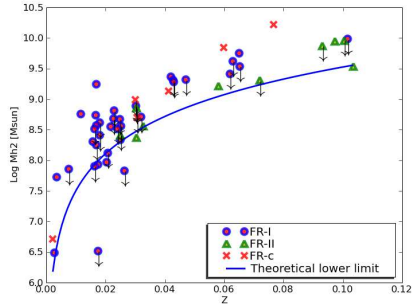}
\caption{$M_{H_2}$ versus z and a lower limit of the Theoretical value for the mass with respect to the distance.}
\label{malmquist}
\end{wrapfigure}
In our sample 69.23\% of the galaxies are FR-I galaxies, 19.23\% are FR-II galaxies and 11.54\% are FR-c. The median value of the masses calculated using ASURV for each subsample shows that the FR-I galaxies ($6.9\times10^7~M_{\odot}$) has less molecular gas mass than the FR-II ($2.9\times10^8~M_{\odot}$) by an order of magnitude. Note that the FR-c galaxies ($5.2\times10^6~M_{\odot}$) are only 6 and for statistical purposes the sample is not big enough.\\
From Figure \ref{FRhisto} it is clear that: (1) Elliptical galaxies do not need much molecular gas mass to host a radio AGN - There is molecular gas mass in the range of 6.3-9.7 (logarithmic values, in units of $M_{\odot}$). (2) There are more FR-I galaxies than FR-II or FR-c in our sample. (3) The FR-I galaxies cover a range of molecular gas mass as large as FR-c and larger than the FR-II. (4) For the median value FR-II galaxies are clearly more massive than the FR-I and FR-c galaxies.\\
The difference in the subsamples can be explained by the Malmquist bias. Figure \ref{malmquist} represents the molecular gas mass vs. z where it is clearly visible that for higher z the galaxies tend to have more molecular gas mass. There is a larger number of FR-II galaxies, compared to FR-I and FR-c galaxies, at higher z implying a large threshold of the upper limit suggesting that this difference could be due to the Malmquist bias.\\

\begin{wrapfigure}[16]{r}{70mm} 
\centering
\includegraphics[scale=0.3]{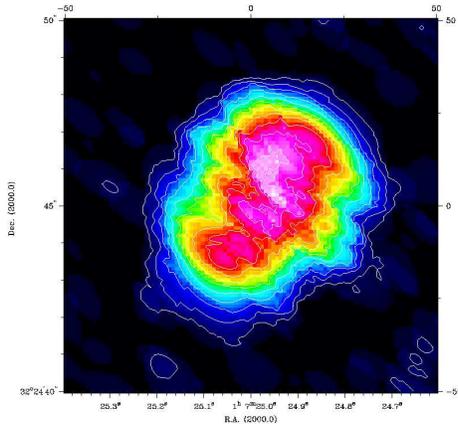}
\caption{CO(1-0) map of the PdBI for 3CR 31 galaxy}
\label{pdbi-3c31}
\end{wrapfigure}

The line ratio between the CO(2-1) and the CO(1-0) transitions is computed with the integrated intensity ratio $I_{CO(2-1)}/I_{CO(1-0)}$ of the CO lines,  where the intensity was measured on one point at the center of each galaxy. As previously noticed by \cite{Lim2000}, the 2 galaxies studied in their paper have a stronger detection in CO(2-1) than CO(1-0). As \cite{Henkel97} noted, this ratio, well over unity, implies a warm ($>$20 K) gas of small column density. This sample has a line ratio well over unity ($>$2). 7 of our  galaxies have been detected in  both frequencies and seven more have been detected only in CO(2-1). From the galaxies detected in both frequencies, only one -NGC 7052- has an integrated emission more intense in the CO(1-0) than in the CO(2-1) line.  The maximum line ratio is found to be 2.48 (for 3CR 31 and 3CR 272.1) and the average value is 2.2 $\pm$ 0.2. Although the beam size is different at the two frequencies, according to \cite{Braine92} it can be compared to the line ratio of about 0.8, the typical value for  spiral galaxies, which also appears closer to the perturbed galaxies, they also suggest that high line ratios are associated with star formation. \\
For those galaxies detected in CO(2-1) emission line, and not detected in CO(1-0) we derived a value for the integrated velocity $I_{CO(1-0)}$ using the ratio of 2.2 calculated here with the galaxies detected in both lines.\\
7 of our galaxies show a double horn line profile typical of a molecular gas disk, representing 36\% of the detected galaxies. We have IRAM-PdBI observations (see Figure \ref{pdbi-3c31}) of 3CR31 in CO(1-0), where a molecular gas disk is clearly visible. \\

\begin{wrapfigure}[17]{r}{75mm} 
\centering
\includegraphics[scale=0.3]{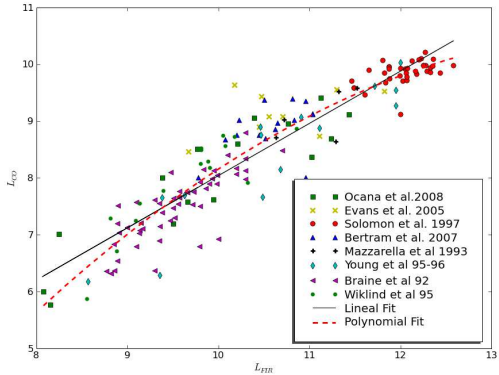}\\ 
\caption{ Our galaxies compare to all the other samples used as comparission. The galaxy sample that fits the best is that of \cite{Solomon97} with ULIRGs.} \label{lum}
\end{wrapfigure}

\section{Dust}

The median value of dust temperature for this sample, using the galaxies detected by IRAS, is 35.8K and median value  of the dust mass is $2.5\times10^6~M_{\odot}$. The dust in this sample is hotter than in spiral galaxies. Possible explanations are that the dust could be heated by an XDR or a high level of star formation. Moreover, the FIR could come directly from the AGN and not from the dust. Finally we noted that the star formation rate is found to be quite low. \\
The $M_{H_2}/M_{dust}$ ratio for this sample, using the galaxies detected for both IRAS and IRAM-30m, is $\sim$300. \\
\newpage
\subsection{FIR vs. CO}
The ratio of $L_{CO}$/$L_{FIR}$ is normally related in spiral galaxies with star formation efficiency. Figure \ref{lum}, is a plot of the galaxies in our sample plus samples used as comparison samples, representing the $L_{FIR}$ vs $L_{CO}$ for the detected galaxies in both FIR and CO luminosities. This plot shows a correlation that appears to be universal. Having a linear fit for all the samples together of $L_{CO}=0.9L_{FIR}-0.9$.\\
Since our data fit well in the plot compared to the other samples, we could argue that elliptical galaxies  should exhibit at least a low level of star formation.
\section{Summary and conclusions}
- We performed a survey of CO(1-0) and CO(2-1) in a sample of nearby radio galaxies chosen only on the basis of their radio continuum.\\
- The detection rate in CO(1-0) and/or CO(2-1) was 38\%.\\
- There is a molecular gas disk in the center for 36\% of the detected galaxies.\\
- Our sample has a high CO(2-1)/CO(1-0) line ratio of 2.2.\\
- The sample has a low molecular gas mass content compared to FIR selected samples.\\
- Radio galaxies do follow the $L_{FIR}-L_{CO}$ correlation.

\bibliographystyle{astron}
\bibliography{ref2}
\end{document}